\documentclass[preprint2]{aastex}

\shorttitle{Top-heavy IMF in UCDs}

\shortauthors{Dabringhausen et al.}

\begin{document}

\title{Low-mass X-ray binaries indicate a top-heavy stellar initial mass function in ultra compact dwarf galaxies}

\author{J\"org Dabringhausen$^1$, Pavel Kroupa$^1$, Jan Pflamm-Altenburg$^1$ and Steffen Mieske$^2$}

\affil{$^1$Argelander-Institut f\"ur Astronomie (AIfA), Auf dem H\"ugel 71, 
  53121 Bonn, Germany}
\affil{$^2$European Southern Observatory, Alonso de Cordova 3107, Vitacura,
  Santiago, Chile}

\email{joedab@astro.uni-bonn.de,pavel@astro.uni-bonn.de, jpflamm@astro.uni-bonn.de,smieske@eso.org}

\begin{abstract}
It has been shown before that the high mass-to-light ratios of ultra compact dwarf galaxies (UCDs) can be explained if their stellar initial mass function (IMF) was top-heavy, i.e. that the IMF was skewed towards high mass stars. In this case, neutron stars and black holes would provide unseen mass in the UCDs. In order to test this scenario with an independent method, we use data on which fraction of UCDs has a bright X-ray source. These X-ray sources are interpreted as low-mass X-ray binaries (LMXBs), i.e. binaries where a neutron star accretes matter from an evolving low-mass star. We find that LMXBs are indeed up to 10 times more frequent in UCDs than expected if the IMF was invariant. The top-heavy IMF required to account for this overabundance is the same as needed to explain the unusually high mass-to-light ratios of UCDs and a top-heavy IMF appears to be the only simultaneous explanation for both findings. Furthermore, we show that the high rate of type~II supernovae (SNII) in the star-burst galaxy Arp~220 suggests a top-heavy IMF in that system. This finding is consistent with the notion that star-burst galaxies are sites where UCDs are likely to be formed and that the IMF of UCDs is top-heavy. It is estimated that the IMF becomes top-heavy whenever the star formation rate per volume surpasses  $0.1 \ {\rm M}_{\odot} \, {\rm yr}^{-1} \, {\rm pc}^{-3}$ in pc-scale regions.
\end{abstract}


\section{Introduction}
\label{sec:intro}

The stellar initial mass function (IMF) quantifies the distribution of stellar masses in a newly born stellar population. Together with the dependency of stellar evolution on stellar mass and metallicity, as well as the rate at which stars are formed in the Universe, the shape of the IMF determines the chemical evolution of the Universe and how its stellar content changes with time. The shape of the IMF also has important implications for the evolution of star clusters. Thus, knowing the shape of the IMF is crucial for a broad variety of astrophysical problems.

Resolved stellar populations in the Milky Way and its satellites support the notion that the IMF does not depend on the conditions under which star formation takes place, but that the stellar masses are distributed according to a single IMF known as the \emph{canonical} IMF \citep{kroupa2001a,kroupa2002a,kumar2008a, bastian2010a}. Ultra compact dwarf galaxies (UCDs) on the other hand provide evidence for the opposite notion, namely that the IMF varies and is top-heavy.

These UCDs are stellar systems that have first been discovered in the Fornax galaxy cluster \citep{hilker1999a}. They have $V$-band luminosities between $10^6$ and some $10^7 \ {\rm L}_{\odot}$, but half-light radii of only about 50 pc or less \citep{drinkwater2003a,mieske2008a}. The confirmed UCDs are at distances where they cannot be resolved into stars with current telescopes, but constrains on their stellar populations can be set by quantities derived from their integrated spectra. One such quantity are the dynamical mass-to-light ($M/L$) ratios of UCDs, i.e mass estimates based on the density profile and the internal velocity dispersion of the UCDs \citep{hasegan2005a,hilker2007a,evstigneeva2007a,mieske2008a}. For a clear majority of the UCDs, the $M/L$ ratios derived from their dynamics are higher than it would be expected if they were pure stellar populations that formed with the canonical IMF \citep{hasegan2005a,dabringhausen2008a,mieske2008a}. This has been taken as evidence for an IMF skewed towards high-mass stars \citep{dabringhausen2009a}, i.e. a top-heavy IMF. The elevated $M/L$ ratios of UCDs would then be explained by a large population of neutron stars and black holes (hereafter called dark remnants), because the age of the UCDs \citep{evstigneeva2007a,chilingarian2008a} implies that all massive stars in them have completed their evolution.

It is plausible that the IMF in UCDs is skewed towards high-mass stars. Molecular clouds massive enough to be the progenitors of UCDs become impenetrable for far-infrared radiation while they 
collapse and become a UCD-type star-cluster. Internal heating of the molecular
cloud leads to a higher Jeans-mass in them preferring the formation
of high-mass stars \citep{murray2009a}. A molecular cloud 
can also be heated  by an external flux of 
highly energetic cosmic rays originating
from a local overabundance of type~II supernovae
increasing the local Jeans-mass \citep{papadopoulos2010a}. 
With the young UCDs being very compact, also crowding of proto-stellar cores and their subsequent 
merging in young UCDs may lead to an overabundance of high-mass stars in them \citep{dabringhausen2010a,weidner2010a}. 

However, the high $M/L$ ratios of UCDs could in principle also be due to non-baryonic dark matter (DM), as was suggested by \citet{goerdt2008a} and \citet{baumgardt2008a}. This is because dark remnants and non-baryonic DM would have the same effect on the $M/L$ ratios of the UCDs, provided that a large enough amount of non-baryonic DM can gather within the UCDs. Note that that non-baryonic DM is an unlikely cause for the high $M/L$ ratios of UCDs, since non-baryonic DM is predicted to gather on rather large scales while UCDs are very compact  \citep{gilmore2007a,murray2009a}. However, in order to exclude this possibility completely, the presence of a sufficient number of dark remnants has to be confirmed independently by a method that does not rely on the fact that dark remnants are non-luminous matter like non-baryonic DM.

Such a method is searching for low-mass X-ray binaries (LMXBs) in UCDs. In these binary systems, a dark remnant and an evolving low-mass star are orbiting around each other. The expanding outer atmosphere of the low-mass companion is accreted by the dark remnant. This matter produces a characteristic X-ray signature. The number of LMXBs depends on the number of NSs and stellar-mass black holes (BHs) and thus on the IMF \citep{verbunt1987a,verbunt2003a}. This implies that stellar systems with a top-heavy IMF can be distinguished from stellar systems with the canonical IMF by an excess of LMXBs.

The formulation of the IMF that is used throughout this paper is introduced in Sec.~(\ref{sec:mass-function}). In Sec.~(\ref{sec:LMXB}), the LMXB-abundance in globular clusters (GCs) and UCDs in dependency of the IMF and this model is compared to observations. The type-II supernova rate in star-bursting galaxies in dependency of the top-heaviness of the IMF is discussed in Sec.~(\ref{sec:SNrate}). It is found in Sec.~(\ref{sec:LMXB}) and Sec.~(\ref{sec:SNrate}), respectively, that the UCDs and the star-bursting galaxy Arp~220 show indications for a top-heavy IMF. This suggests that the star formation rate per volume is perhaps the parameter that determines whether the IMF in that volume becomes top-heavy, as is argued in Sec.~(\ref{sec:discussion}). Conclusions are given in Sec.~(\ref{sec:conclusion}).

\section{The initial stellar mass function}
\label{sec:mass-function}

A varying IMF can be formulated as
\begin{equation}
\xi (m) =k \, k_i \, m^{-\alpha_i},
\label{eq:IMF}
\end{equation}
with 
\begin{eqnarray}
\nonumber \alpha_1 = 1.3,  & \qquad & 0.1  \le  \frac{m}{\rm{M}_{\odot}} < 0.5,\\
\nonumber \alpha_2 = 2.3,  & \qquad & 0.5  \le  \frac{m}{\rm{M}_{\odot}} < m_{\rm tr}, \\
\nonumber \alpha_3 \, \in \, \mathbb{R}, & \qquad & m_{\rm tr} \le  \frac{m}{\rm{M}_{\odot}} \le m_{\rm max},
\end{eqnarray}
where $m$ is the initial stellar mass, the factors $k_i$ ensure that the IMF is continuous where the power changes and $k$ is a normalization constant. $\xi(m)$ equals 0 if $m<0.1 \, {\rm M}_{\odot}$ or $m> m_{\rm max}$, where $m_{\rm max}$ is a function of the star-cluster mass \citep{weidner2006a,weidner2010b} and $m_{\rm tr}$ is the stellar mass at which the IMF begins to deviate from the canonical IMF. For $m_{\rm tr} = 1 \, {\rm M}_{\odot}$, the formulation of the IMF used here is identical with the one used in \citet{dabringhausen2009a}, so that results found here for this choice of $m_{\rm tr}$ can be compared to results in \citet{dabringhausen2009a}. For $\alpha_3 = \alpha_2 = 2.3$, Equation~(\ref{eq:IMF}) is the canonical IMF \citep{kroupa2001a,kroupa2002a}. For $\alpha_3 < 2.3$, the IMF is top-heavy, implying more intermediate-mass stars and in particular more high-mass stars.

In the mass range of UCDs, $m_{\rm max}$ is not set by the mass of the stellar system, but by the observed mass limit for stars,  $m_{\rm max *}$. Thus, $m_{\rm max}=m_{\rm max *}$ for all UCDs. The actual value of $m_{\rm max *}$ is, however, rather uncertain: Estimates range from the canonical value $m_{\rm max *} \approx 150 \ {\rm M}_{\odot}$ \citep{weidner2004a,oey2005a} to $m_{\rm max *} \approx 300 \ {\rm M}_{\odot}$ (\citealt{crowther2010a}, but see \citealt{banerjee2011a}). In this paper,  $m_{\rm max *} = 150 \ {\rm M}_{\odot}$ is assumed, but note that assuming $m_{\rm max *} = 300 \ {\rm M}_{\odot}$ instead would have little effect on the results reported here (see Section~\ref{sec:results} and Figure~\ref{fig:comp}).

In the case of GCs and UCDs with LMXBs (see Section~\ref{sec:LMXB}), the observed luminosity, $L$, is known to originate from stars with masses $m\lesssim 1 \ {\rm M}_{\odot}$. This is because their stellar populations are old \citep{evstigneeva2007a,chilingarian2008a} and the more massive stars have already completed their evolution. Being fixed by observations, $L$ should however not be changed when the IMF is varied. For the IMF given by Equation~(\ref{eq:IMF}), this can be achieved by finding $k$ from the condition
\begin{equation}
\int^{m_{\rm max *}}_{0.1\ {\rm M}_{\odot}} \xi_{\rm can} (m)m \, dm=1 \, \rm{M}_{\odot},
\label{eq:norm1}
\end{equation}
where $\xi_{\rm can}$ is the canonical IMF, i.e. $\alpha_3=2.3$. With this normalization, the number density of stars with $m<1 \ {\rm M}_{\odot}$ is the same for all values of $\alpha_3$, since the normalization is set by the canonical IMF and is therefore not affected by variations of $\alpha_3$.

In the case of the SN-rate of Arp~220 (see Section~\ref{sec:SNrate}), the light used to estimate the star formation rate (SFR), i.e. the mass of the material converted into stars per time-unit, originates from stars over the whole range of stellar masses. With the SFR thereby given, we then normalize the IMF such that the SFR remains constant when the IMF is varied. For the IMF given by Equation~(\ref{eq:IMF}), this can be achieved by finding $k$ from the condition
\begin{equation}
\int^{m_{\rm max *}}_{0.1\ {\rm M}_{\odot}} \xi (m)m \, dm=1 \, \rm{M}_{\odot}.
\label{eq:norm2}
\end{equation}
With this normalization, the number density of stars with $m<1 \ {\rm M}_{\odot}$ decreases with decreasing values of $\alpha_3$, i.e. with increasing top-heaviness of the IMF.

Stellar evolution and dynamical evolution turn the IMF of a star cluster into a (time-dependent) mass function of stars and stellar remnants; the star and stellar remnant mass function, SRMF. For a single-age stellar population, the connection between the IMF and the SRMF can be quantified by an initial-to-final mass relation for stars, $m_{\rm rem}$, which can be written as
\begin{equation}
 m_{\rm rem} =  \left\{
 \begin{array}{ll}
\displaystyle \frac{\displaystyle m}{\displaystyle {\rm M}_{\odot}}, & \quad \displaystyle \frac{\displaystyle m}{\displaystyle {\rm M}_{\odot}} < \displaystyle \frac{\displaystyle m_{\rm{to}}}{\displaystyle {\rm M}_{\odot}}, \\
 & \\[-6 pt]
 0.109 \, \displaystyle \frac{\displaystyle m}{\displaystyle {\rm M}_{\odot}}+0.394, & \quad \displaystyle \frac{\displaystyle m_{\rm{to}}}{\displaystyle {\rm M}_{\odot}} \le \displaystyle \frac{\displaystyle m}{\displaystyle {\rm M}_{\odot}} < 8, \\
 & \\[-6 pt]
 1.35, & \quad 8  \leq  \displaystyle \frac{\displaystyle m}{\displaystyle{\rm M}_{\odot}} < 25, \\
 & \\[-6 pt]
  0.1\displaystyle \frac{\displaystyle m}{\displaystyle {\rm M}_{\odot}}, & \quad 25  \leq \displaystyle \frac{\displaystyle m}{\displaystyle {\rm M}_{\odot}} \le m_{\rm{max *}},
  \end{array}
  \right.
\label{eqMrem1}
\end{equation}
where $m_{\rm to}$ is the mass at which stars evolve away from the main sequence at a given age \citep{dabringhausen2009a}. UCDs typically have ages of $\approx$ 10 Gyr \citep{evstigneeva2007a,chilingarian2008a}, which implies $m_{\rm to} \approx 1 \ \rm{M}_{\odot}$ for them. In the present paper, Equation~(\ref{eqMrem1}) is used to calculate how the mass of a modeled UCD depends on the variation of its IMF (see Section~\ref{sec:variableIMF}).

Note that Equation~(\ref{eqMrem1}) reflects the evolution of single stars. In a binary system, the initial mass of a star that evolves into a black hole is expected to be higher, so that stars with masses up to 40 ${\rm M}_{\odot}$ may become NSs instead of BHs (cf. \citealt{brown2001a}). It is however of minor importance in this paper whether a massive remnant is a NS or a BH. Both kinds of objects can become bright X-ray sources by accreting matter from a companion star and BHs in such binary systems are actually detected by excluding that they are NSs due to their mass \citep{casares2007a}. Also the total mass of a GC or UCD is not strongly affected by the mass-limit between NSs and BHs. Using Equation~(ref{eq:IMF}) with $m_{\rm tr} = 1 \ {\rm M}_{\odot}$ and Equation~(\ref{eqMrem1}) with $m_{to}=1 \ {\rm M}_{\odot}$, the total mass of NSs and BHs is 4.2 per cent of the total mass of the stellar system for $\alpha_3=2.3$ (canonical IMF) and 79.9 per cent for $\alpha_3=1$. These numbers are altered to 3.8 per cent of the total mass of the stellar system for $\alpha_3=2.3$ and 75.0 per cent for $\alpha_3=1$ if the transition from NSs to BHs is shifted from $25 \ {\rm M}_{\odot}$ to $40 \ {\rm M}_{\odot}$.

\section{The LMXB-abundance in GCs and UCDs}
\label{sec:LMXB}

\subsection{Some properties of GCs and UCDs}
\label{sec:properties}

\begin{figure}[t]
\centering
\includegraphics[scale=0.7]{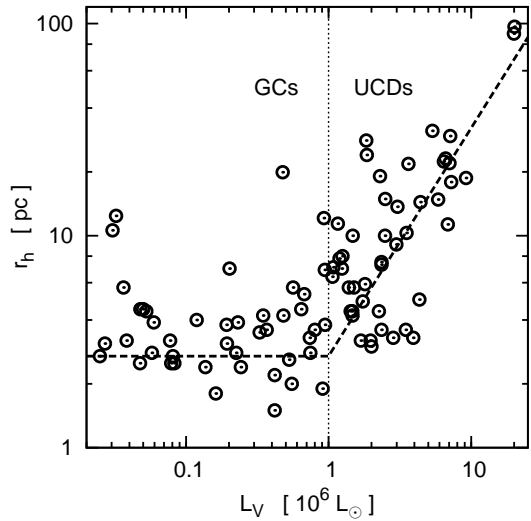}
\caption{\label{fig:MRad} The effective half-light radii, $r_{\rm h}$, of GCs and UCDs. The circles show the sample of individual GCs and UCDs from the compilation of \citet{mieske2008a}. The dashed line is an estimate of the average $r_{\rm h}$ of GCs and UCDs (cf. Equations~\ref{eq:rh1} and~\ref{eq:rh2}). The vertical dotted line sets the limit between objects that are considered as GCs and objects that are considered as UCDs. Note that the average $r_{\rm h}$ indicated for GCs by the dashed line is lower than the average $r_{\rm h}$ of the GCs shown in this figure. This is because the GCs shown here are mostly GCs of the Milky Way while the dashed line corresponds to the average $r_{\rm h}$ of GCs in the Virgo-cluster. The GCs in the Virgo cluster tend to be more compact than those around the Milky Way.}
\end{figure}

\begin{figure}[t]
\centering
\includegraphics[scale=0.7]{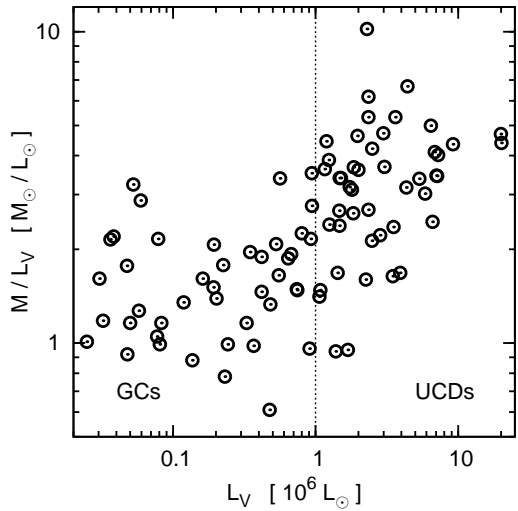}
\caption{\label{fig:ML} The mass-to-light ratios (M/L ratios) of GCs and UCDs. The circles show the sample of individual GCs and UCDs from the compilation of \citet{mieske2008a}. The vertical dotted line sets the limit between objects that are considered as GCs and objects that are considered as UCDs.}
\end{figure}

For a number of GCs and UCDs, data \citep{mieske2008a} on $V$-band luminosity ($L_V$), dynamical mass ($M_{\rm dyn}$) and effective half-light radius ($r_{\rm h}$) are available. These data suggest a transition at  $L_V \approx 10^6 \ {\rm L}_{\odot}$, since the $r_{\rm h}$ and dynamical $M/L$ ratios of objects with $L_V < 10^6 \ {\rm L}_{\odot}$ appear to be independent of $L_V$, in contrast to objects with $L_V > 10^6 \ {\rm L}_{\odot}$ (see Figures~\ref{fig:MRad} and~\ref{fig:ML}). This motivates to consider the objects with $L_V < 10^6 \ {\rm L}_{\odot}$ as GCs and those with  $L_V \ge 10^6 \ {\rm L}_{\odot}$ as UCDs, even though stellar systems close to this transition could be assigned to either one of these classes \citep{mieske2008a}.

\begin{figure}[t]
\centering
\includegraphics[scale=0.7]{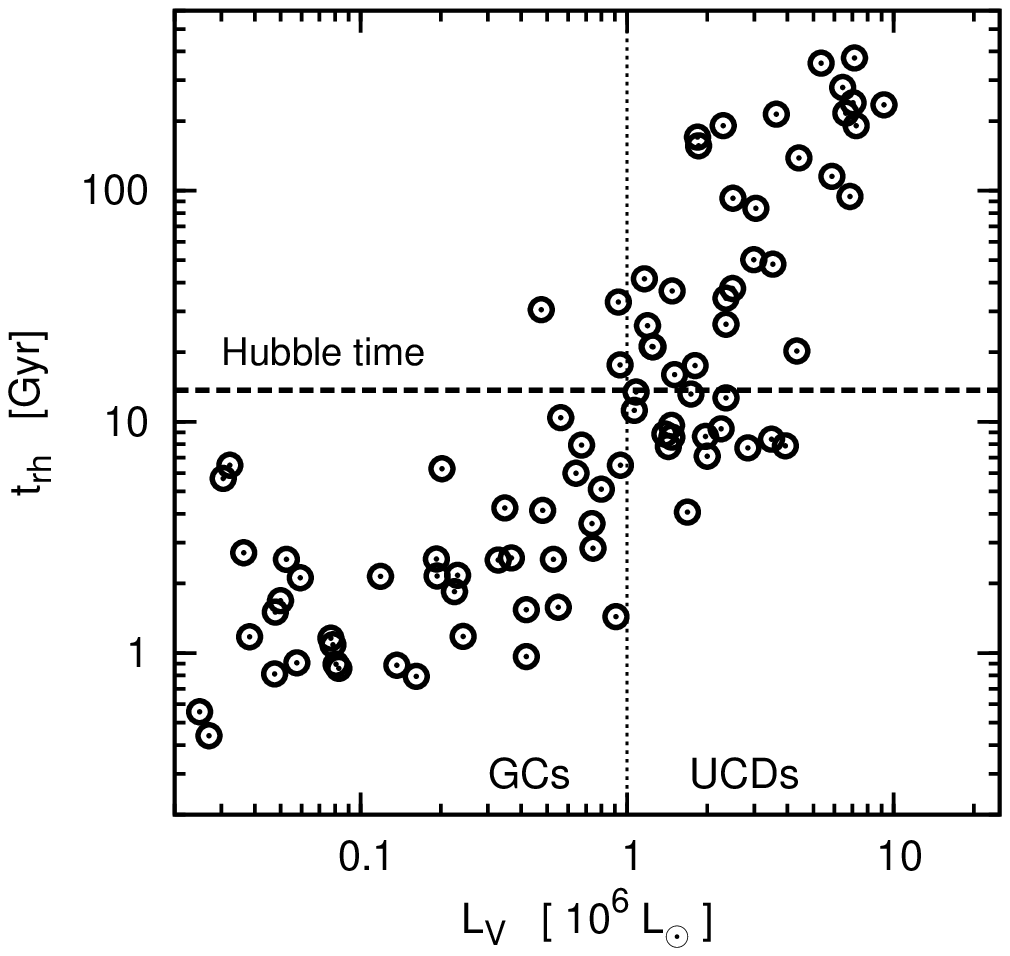}
\caption{\label{fig:trelax} The median two-body relaxation times, $t_{\rm rh}$,  of GCs and UCDs. The circles show the sample of individual GCs and UCDs from the compilation of \citet{mieske2008a}. The dashed horizontal line indicates the age of the Universe, $\tau_{\rm H}$, according to the $\Lambda$CDM-model. The vertical dotted line sets the limit between objects that are considered as GCs and objects that are considered as UCDs. Note that $t_{\rm rh} \gtrsim \tau_{\rm H}$ for UCDs. Thus, UCDs can be considered dynamically unevolved \citep{dabringhausen2008a} and they may therefore be considered as galaxies from a stellar dynamical point of view \citep{forbes2011a}.}
\end{figure}

Knowing $r_{\rm h}$ and $M_{\rm dyn}$ of a stellar system allows to estimate its median two-body relaxation time \citep{spitzer1987b}, using
\begin{equation}
t_{\rm rh}=\frac{0.234}{\log_{10}(M_{\rm dyn}/{\rm M}_{\odot})} \times \sqrt{\frac{M_{\rm dyn} \, r_{\rm h}^3}{G}},
\label{eq:trelax}
\end{equation}
where $G$ is the gravitational constant \citep{dabringhausen2008a}. The significance of $t_{\rm rh}$ lies in the fact that it sets the time-scale on which the structure of a self-bound stellar system is changed by the process of energy equipartition. If $\tau \gtrsim t_{\rm rh}$ holds for a stellar system with $\tau$ being its age, it can be considered nearly unaffected by dynamical evolution and is thus only subject to stellar evolution. This is the case for UCDs, as $t_{\rm rh} \gtrsim \tau_{\rm H}$ is valid for them, where $\tau_{\rm H}$ is the age of the Universe suggested by the $\Lambda$CDM-model (see Figure~\ref{fig:trelax}). Thus, the properties of UCDs can be calculated from their IMF while considering the effects of stellar evolution, but without accounting for the effects of dynamical evolution. This means in particular that the SRMF of UCDs can be calculated from their IMF and Equation~(\ref{eqMrem1}). Note that GCs, on the other hand, \emph{are} subject to dynamical evolution, since their ages, $\tau_{\rm GC}$ are also similar to $\tau_{\rm H}$ and thus $\tau_{\rm GC} > t_{\rm rh}$.

The data \citep{mieske2008a} on $L_V$ and $r_{\rm h}$ of individual GCs in the MW and in Centaurus A and UCDs in the Virgo-cluster are also useful for estimating an average $r_{\rm h}$, $\overline{r}_{\rm h}$, as a function of $L_V$. GCs over the luminosity range from $10^4 \ {\rm L}_{\odot}$ to $10^6 \ {\rm L}_{\odot}$ do not show a luminosity-radius trend \citep{mclaughlin2000a,jordan2005a}. The logarithmic average $r_{\rm h}$ of GC is
\begin{equation}
\log_{10} \left(\frac{\overline{r}_{\rm h}}{\rm pc}\right)= 0.4314
\label{eq:rh1}
\end{equation}
\citep{jordan2005a}. Performing a linear least-squares fit to data in \citet{mieske2008a} on UCDs in the Virgo cluster leads to
\begin{equation}
\log_{10} \left(\frac{\overline{r}_{\rm h}}{\rm pc}\right)=1.076\log_{10} \left(\frac{L_V}{10^6 \, L_{V,\odot}}\right)+0.4314.
\label{eq:rh2}
\end{equation}
Note that equality between Equations~(\ref{eq:rh1}) and~(\ref{eq:rh2}) at $L_V = 10^6 \ {\rm L}_{\odot}$ was imposed as a secondary condition on the fit of Equation~(\ref{eq:rh2}) to the data. This secondary condition reflects the fact that the $r_{\rm h}$ of GCs are indistinguishable from those of UCDs at $L\approx10^6 \ {\rm L}_{\odot}$ (see Figures~\ref{fig:MRad} and \ref{fig:ML}).

\subsection{Modeling the LMXB-abundance in GCs and UCDs}
\label{sec:enc-rate}

\subsubsection{The origin of LMXBs in GCs and UCDs}
\label{sec:origin}

Tight binaries consisting of a dark remnant and a low-mass companion can have in principle two different origins: 
\begin{enumerate}
\item They can be primordial. In this case a tight binary of a high-mass star and a low-mass star have formed already in the star forming event. The high-mass star explodes in a supernova after a few million years leaving behind a dark remnant which can remain bound to its low-mass companion.
\item They have formed through encounters. GCs and UCDs are regions of enormously high  stellar density ranging from $10 \ {\rm M}_{\odot} \, {\rm pc}^{-3}$ to $10^4 \ {\rm M}_{\odot} \, {\rm pc}^{-3}$ \citep{dabringhausen2008a}. Encounters between dark remnants and low-mass stars are therefore frequent and can lead to the formation of LMXBs due to tidal capture \citep{verbunt1987a,verbunt2003a}.
\end{enumerate}
As these formation mechanisms are quite different it is expected that both processes would contribute differently to the LMXB content in GCs and UCDs.

There are however strong arguments against a significant contribution from primordial binaries to the LMXB content of GCs and UCDs:
\begin{enumerate}
\item The number of LMXBs in GCs is strongly correlated with the encounter rate and thus clearly linked to it \citep{jordan2005a,sivakoff2007a}.
\item There are several hundred times more LMXBs per unit mass in GCs than in the Galactic field \citep{verbunt1987a}. The LMXBs in the Galactic field are LMXBs that probably evolved from primordial binaries, since they are in a low-density environment where encounters play no role and most probably formed in star clusters from which they were subsequently ejected. The strong excess of LMXBs in GCs therefore suggests that most LMXBs in GCs form through encounters \citep{verbunt1987a}.
\end{enumerate}

The number of encounters relevant for the creation of LMXBs, i.e. encounters where a NS can capture a low-mass star \citep{verbunt1987a}, can be written as
\begin{equation}
\Gamma \propto \frac{n_{\rm ns} \, n_{\rm s} \, r_{\rm c}^3}{\sigma},
\label{eq:Gamma3}
\end{equation}
where  $n_{\rm ns}$ is the number density of NSs, $n_{\rm s}$ is the number density of potential low-mass companion stars, $r_{\rm c}$ is the core radius of the stellar system and $\sigma$ is the velocity dispersion \citep{verbunt2003a}. The potential companions to a NS in a bright LMXB are stars that come from a rather narrow mass range where stars of a given age leave the main-sequence. At this stage of their evolution, the stars expand rapidly, which makes a high accretion rate on the NS possible, which in turn leads to a high X-ray luminosity.

A more recent study by \citet{ivanova2008a} revealed that tidal capture is not the only dynamical process relevant for the formation of LMXBs. Other dynamical processes like direct collisions between NSs and red giant stars or interactions between stars and existing binaries also play a role and can actually be even more important than tidal captures. However, this does not change the observational finding that the number of LMXBs in GCs scales with $\Gamma$ (e.g. \citealt{jordan2005a}). Therefore, $\Gamma$ seems to be an adequate measure for the stellar dynamical processes that produce LMXBs in general. It is moreover argued in \citet{ivanova2008a} that primordial binaries only make a small contribution to the total population of LMXBs in old GCs. 

There are thus strong observational and theoretical motivations for the usage of $\Gamma$ as a measure for how many LMXBs are expected in GCs and UCDs.

\subsubsection{The encounter rate in GCs and UCDs for an invariant SRMF}
\label{sec:constantIMF}

If only a \emph{single}, invariant mass function for stars and stellar remnants (SRMF) is considered for all stellar systems, then
\begin{equation}
n_{\rm ns} \propto n_{\rm s} \propto \rho_0
\label{eq:rho1}
\end{equation}
holds, where $\rho_0$ is the central mass density. Equation~(\ref{eq:Gamma3}) can then be rewritten as
\begin{equation}
\Gamma \propto \frac{\rho_0^2 \, r_{\rm c}^3}{\sigma}
\label{eq:Gamma4}
\end{equation}
by using Equation~(\ref{eq:rho1}).

In order to link the theory on LMXB-formation to the optical properties of observed stellar systems, it is in the following assumed that the mass density of a stellar system follows its luminosity density. The structural parameters derived from the distribution of the light in the stellar system can then be translated directly into statements on the distribution of its mass, i.e. quantities that determine the dynamics of the stellar system.

However, $r_{\rm c}$ is difficult to measure for GCs and UCDs at the distance of the Virgo cluster, as these stellar systems are barely resolved with current instruments. The projected half-light radius $r_{\rm h}$ (and thus the half-mass radius under the assumption that mass follows light) is larger and therefore less difficult to retrieve from the data. For practical purposes, it is therefore useful to assume
\begin{equation}
r_{\rm c} \propto r_{\rm h}
\label{eq:rc}
\end{equation}
and
\begin{equation}
\rho_0 \propto  \frac{M}{r_{\rm h}^3},
\label{eq:rho2}
\end{equation}
where $M$ is the mass of the stellar system. The King profile \citep{king1962a} with its three independent parameters (core radius, tidal radius and central density), is thereby simplified to a density profile with only two independent parameters (half-mass radius and mass). The underlying assumptions are not necessarily true, and indeed, not fulfilled for GCs in the Milky  Way since \citet{mclaughlin2000a} finds that more luminous GCs tend to be more concentrated. However, regarding the conclusions on how the presence of bright LMXBs is connected to the optical properties of GCs in the Virgo cluster, these assumptions are unproblematic. Using the same concentration for all GCs in their sample \citet{sivakoff2007a} find that they essentially come to the same results as \citet{jordan2004a}, who use an individual estimate for the concentration of each GC in their sample. 

When dealing with UCDs, replacing $r_{\rm c}$ with $r_{\rm h}$ is even advantageous. The time-scale on which the NSs gather at the centre of the UCD is given as
\begin{equation}
t_{\rm seg}=\frac{\overline{m}}{m_{\rm ns}} \, t_{\rm cc},
\label{eq:seg}
\end{equation}
where $\overline{m} \approx 0.5 \ {\rm M}_{\odot}$ is the mean mass of stars, $m_{\rm ns} \approx 1.35 \ {\rm M}_{\odot}$ is the mass of neutron stars and $t_{\rm cc}$ is the core-collapse time of the UCD without the NSs \citep{spitzer1987b,banerjee2010a}. If a Plummer sphere \citep{plummer1911a} is used as an approximation for the density profile of a stellar system, $t_{\rm cc} \approx 15 \ t_{\rm rh}$ holds \citep{baumgardt2002a}. With $t_{\rm rh}$ being of the order of a Hubble time for UCDs, Equation~(\ref{eq:seg}) implies that the distribution of NSs in UCDs still follows the initial distribution of their progenitors.  The volume relevant for the formation of LMXBs in a UCD is therefore better measured by $r_{\rm h}$ than by $r_{\rm c}$, provided its stellar population did not \emph{form} mass-segregated. This is because $r_{\rm h}$ represents the size of the whole UCD, whereas $r_{\rm c}$ represents the size of its centre.

Thus, using Equations~(\ref{eq:rc}) and~(\ref{eq:rho2}), Equation~(\ref{eq:Gamma4}) can be transformed into
\begin{equation}
\Gamma_{\rm h} \propto \frac{M^2 }{r_{\rm h}^3 \, \sigma}.
\label{eq:Gamma5}
\end{equation}
If the stellar system is also in virial equilibrium, 
\begin{equation}
\sigma \propto \rho_0^{0.5} \, r_{\rm h} \propto \frac{M^{0.5}}{r_{\rm h}^{0.5}}
\label{eq:vir}
\end{equation}
holds. In this case,
\begin{equation}
\Gamma_{\rm h} \propto \frac{M^{1.5}}{r_{\rm h}^{2.5}}
\label{eq:Gamma6}
\end{equation}
follows from Equations~(\ref{eq:Gamma5}) and~(\ref{eq:vir}). In contrast to Equation~(\ref{eq:Gamma3}), Equation~(\ref{eq:Gamma6}) has only two variables ($M, r_{\rm h}$) instead of four ($n_{\rm s}, n_{\rm ns}, M, r_{\rm c}$).

A further variable can be eliminated by replacing individual values for $r_{\rm h}$ by luminosity-dependent estimates for $r_{\rm h}$, such as Equations~(\ref{eq:rh1}) and~(\ref{eq:rh2}), and noting that the same SRMF for all stellar systems in question implies $M \propto L_V$ for them. This leads to 
\begin{equation}
\overline{\Gamma}_{\rm h} \propto \frac{L_V^{1.5}}{\overline{r}_{\rm h}^{2.5}}, 
\label{eq:Gamma7}
\end{equation}
or, more explicitly by using Equations~(\ref{eq:rh1}), and~(\ref{eq:rh2}), respectively,
\begin{equation}
\log_{10} (\overline{\Gamma}_{\rm h}) = 1.5\log_{10} \left(\frac{L_V}{10^6 \, {\rm L}_{\odot}}\right)+A
\label{eq:GammaGC}
\end{equation}
for GCs (i.e. $L_V<10^6 \ {\rm L}_{\odot}$), and
\begin{equation}
\log_{10} (\overline{\Gamma}_{\rm h}) = -1.190\log_{10} \left(\frac{L_V}{10^6 \, {\rm L}_{\odot}}\right)+A
\label{eq:GammaUCD}
\end{equation}
for UCDs (i.e. $L_V \ge 10^6 \ {\rm L}_{\odot}$). The constant $A$ is the same in Equations~(\ref{eq:GammaGC}) and~(\ref{eq:GammaUCD}). Note that the transition between Equations~(\ref{eq:GammaGC}) and~(\ref{eq:GammaUCD}) is continuous due to the continuity of $\overline{r}_{\rm h}$ at $L_V=10^6 \ {\rm L}_{\odot}$. 

\subsubsection{Detecting a variable SRMF with LMXBs}
\label{sec:variableIMF}

For investigating how $\Gamma_{\rm h}$ depends on the IMF, it is useful to consider the ratio between $\Gamma_{\rm h}$ as a function of $\alpha_3$ and the $\Gamma_{\rm h}$ implied by some reference IMF. This has the advantage that factors, which do not depend on the IMF, cancel. The reference IMF is the canonical IMF in this paper; a choice that is motivated with the lack of dynamical evolution in UCDs (cf. Section~\ref{sec:properties}). Using Equation~(\ref{eq:rho2}) thus leads to
\begin{equation}
\frac{\Gamma_{\rm h}(\alpha_3)}{\Gamma_{\rm h}(\alpha_3=2.3)}=\\
\frac{n_{\rm ns}(\alpha_3)}{n_{\rm ns}(\alpha_3=2.3)} \sqrt{\frac{M(\alpha_3=2.3)}{M(\alpha_3)}},
\label{eq:ratio2}
\end{equation}
if it also assumed that the IMF varies only for stars with $m>m_{\rm to}$, so that also $n_{\rm s}$ is constant. By this last assumption, the luminosity of the UCDs, which is given by observations, stays constant when the IMF of the UCDs is varied.  The right side of Equation~(\ref{eq:ratio2}) can be calculated if the IMF is specified. In particular,
\begin{equation}
\frac{n_{\rm ns}(\alpha_3)}{n_{\rm ns}(\alpha_3=2.3)}=\frac{\int^{m_{\rm max *}}_{8 \ {\rm M}_{\odot}} \xi (m) \, dm}{\int^{m_{\rm max *}}_{8 \ {\rm M}_{\odot}} \xi_{\rm can} (m) \, dm},
\end{equation}
and
\begin{equation}
\frac{M(\alpha_3=2.3)}{M(\alpha_3)}=\frac{\int^{m_{\rm max *}}_{0.1\ {\rm M}_{\odot}} m_{\rm rem}(m) \xi_{\rm can}(m) \, dm}{\int^{m_{\rm max *}}_{0.1\ {\rm M}_{\odot}} m_{\rm rem}(m) \xi(m) \, dm},
\end{equation}
where the IMF is normalised using Equation~(\ref{eq:norm1}), $\xi_{\rm can}$ is the canonical IMF and $m_{\rm rem}(m)$ is given by Equation~(\ref{eqMrem1}). Thus, Equation~(\ref{eq:ratio2}) quantifies how $\Gamma_{\rm h}$ changes in a stellar system (normalized with $\Gamma_{\rm h}$ for the canonical IMF) if the number of dark remnants and therefore the mass of the stellar system are changed, while its characteristic radius and the number of stars are kept constant.

A difficulty is that $\Gamma_{\rm h}$ of a stellar system cannot be measured directly. However, the actual $\Gamma_{\rm h}$ of  a GC or a UCD scales with the rate at which LMXBs are created (see Sec.~\ref{sec:origin}), which is proportional to the probability $P$ to form an LMXB above a certain brightness limit in a given time. If a sample of GCs or UCDs in a certain luminosity interval is given, a useful estimator for the average $P$ of these GCs or UCDs is the fraction $f_{\rm LMXB}$ of them that have an LMXB above the brightness limit defined by the sensitivity of a given set of observations. Thus,
\begin{equation}
\label{eq:prob}
f_{\rm LMXB} \propto P \propto \Gamma_{\rm h}^{\gamma},
\end{equation}
where the exponent $\gamma$ accounts for the claims that the LMXB-frequency in GCs and UCDs may not be directly proportional to $\Gamma$ or $\Gamma_{\rm h}$, but to some power of $\Gamma$ or $\Gamma_{\rm h}$ (cf. \citealt{jordan2004a,sivakoff2007a}).

If the SRMF of UCDs is indeed independent of luminosity, the $f_{\rm LMXB}$ of UCDs in different $L_V$ intervals should all roughly coincide with the prediction from Equation~(\ref{eq:GammaUCD}) for an appropriate choice of the constant $A$. If however the $f_{\rm LMXB}$ of at least one $L_V$ interval is inconsistent with Equation~(\ref{eq:GammaUCD}) for any choice of $A$, then this would be evidence for the SRMF changing with the luminosity of the UCDs. This would imply that the IMF of the UCDs changes with luminosity, since the SRMF of UCDs is solely determined by stellar evolution, i.e. a process that does not depend on the size of the system (see Section~\ref{sec:properties}). Note that the actual value of $A$  in Equations~(\ref{eq:GammaGC}) and~(\ref{eq:GammaUCD}) has no implications for the physical properties of the observed stellar systems: For a given sample of GCs and UCDs, $A$ depends on the detection limit for an X-ray source or an arbitrarily chosen brightness limit above the detection limit.

\subsubsection{Data on the LMXB-frequency in GCs and UCDs}
\label{sec:data}

In order to search for a dependency of the IMF in UCDs on their luminosity, we use data published in the upper left panel of figure~(6) in \citet{sivakoff2007a}. These data provide the fraction
of globular clusters and UCDs, $f_{\rm LMXB}$, hosting an LMXB 
in a given total $z$-band magnitude interval.

The results of \citet{sivakoff2007a} were obtained by combining two sets of data. 

First, HST images of 11 elliptical galaxies in the Virgo Cluster were used, see Table~1 in \citet{sivakoff2007a}. Ten of them are the brightest galaxies observed in the course of the ACS Virgo Cluster Survey \citep{cote2004a}. The eleventh one (NGC 4697) is a similarly bright galaxy that was observed by \citet{sivakoff2007a} with nearly the same observational setup as in the ACS VIrgo Cluster Survey. Using the obtained images, a large number of accompanying GCs and UCDs was identified around each of these galaxies.

Second, \citet{sivakoff2007a} used archival Chandra Observatory X-ray observations of the same galaxies. The setup for the X-ray observations  varied widely from galaxy to galaxy, see Table~2 in \citet{sivakoff2007a}, which could in principle be problematic.

\citet{sivakoff2007a} find however that the global properties of GCs and UCDs which contain a LMXB are largely unaffected by the varying detection limits for X-ray sources. Also note that the LMXB-frequencies in GCs are well explained by the encounter rates in them (see Section~\ref{sec:results}), despite the different detection limits for X-ray sources. This suggests that the encounter rate is indeed a good measure for the rate at which LMXBs of any X-ray luminosity are created. We therefore assume that a large number of GCs and UCDs with an X-ray source is indeed an indicator for a large number of dark remnants in them.

The size of the $z$-band magnitude intervals in \citet{sivakoff2007a} is chosen such that each of them contains 27 GCs or UCDs with a detected LMXB. This corresponds to a total of at least 100 GCs or UCDs in each of these intervals, since $f_{\rm LMXB} \lesssim 0.2$ in all of them. Thus, $f_{\rm LMXB}$ can be taken as a reliable estimator for the average $P$ to form an LMXB in a GC or a UCD in a given $z$-band magnitude interval.

For comparing the data on the LMXBs in GCs and UCDs from \citet{sivakoff2007a} to the prediction for the LMXB-frequency in GCs and UCDs formulated in Equations~(\ref{eq:GammaGC}) and~(\ref{eq:GammaUCD}), $z$-band magnitudes have to be converted into $L_V$. For this purpose, $z$-band luminosities are calculated from $z$-band magnitudes with 
\begin{equation}
L_z=10^{-0.4(M_z-4.51)} {\rm L}_{\odot,z},
\label{eq:zLum}
\end{equation}
where $M_z$ is the absolute $z$-band magnitude and $L_z$ is $z$-band luminosity in Solar units (cf. Equation~1 in \citealt{sivakoff2007a}). Now note that the $z$-band $M/L$ ratio of  GCs in the Virgo-cluster are all close to $\approx 1.5 {\rm M}_{\odot}/L_{\odot,z}$ \citep{sivakoff2007a}, which is essentially identical to the average $V$-band $M/L$ ratio of the GCs in the Milky Way in Solar units \citep{mclaughlin2000a}. This implies that $z$-band and 
$V$-band luminosities of GCs are approximately identical in Solar units. We therefore assume $L_{V}/L_{\odot,V}=L_z/L_{\odot,z}$ in this paper.

\begin{figure*}[t]
\centering
  \includegraphics[scale=0.7]{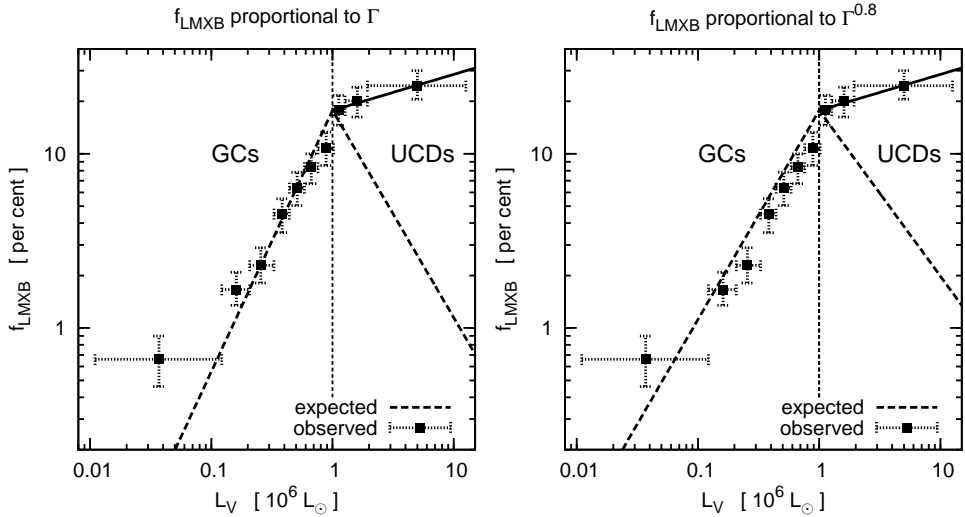}
  \caption{\label{lmbx-lv} The observed LMXB-frequency of GGs and UCDs in comparison to expected frequencies if the IMF was canonical. Plotted are the observed frequencies (squares) of GCs ($L_V < 10^6 \ L_{\odot,V}$) and UCDs ($L_V \ge 10^6 \ {\rm L}_{\odot,V}$) with LMXBs, $f_{\rm LMXB}$, in 
    the Virgo galaxy cluster as a function of the V-band luminosity, $L_V$.
    Each data point contains 27 objects showing the LMXB signal. The data points are identical with the data points in the upper left panel of figure~(6) in \citet{sivakoff2007a}, except for a rescaling of $z$-band magnitudes to $V$-band luminosities.
    The three brightest data are based on $\approx$400 UCDs, i.e 
    $\approx$135 UCDs per bin.   
    The dashed line shows the theoretically expected LMXB-frequency
    for an invariant canonical IMF with index $\alpha_3=2.3$ assuming $f_{\rm LMXB} \propto \overline{\Gamma}_{\rm h}$ (left panel) and $f_{\rm LMXB} \propto \overline{\Gamma}_{\rm h}^{0.8}$ (right panel). 
    In either case, the theoretically expected LMXB-frequency is significantly too low for UCDs, while for GCs the theoretically expected LMXB-frequency matches the observed LMXB-frequency.
    The solid line is a fit through the UCD-regime (above $10^6 \ {\rm L}_{\odot,V}$).
    From it is derived the variation with luminosity of the IMF index 
    $\alpha_3$ such that this new model, based on a variable IMF, 
    accounts for the observed $f_{\rm LMXB}$ for $L_V > 10^6 \ {\rm L}_{\odot,V}$.       
  } 
\end{figure*}

The data from figure~(6) in \citet{sivakoff2007a} is shown in Figure~\ref{lmbx-lv} with the $z$-band magnitude intervals from \citet{sivakoff2007a} converted into $L_V$ intervals. Three of these intervals are at luminosities $L_V>10^6 \ {\rm L}_{\odot}$, so that the objects in them are UCDs (cf. Section~\ref{sec:properties}). As the size of the intervals is chosen such that each of them contains 27 objects with an LMXB, 81 UCDs with an LMXB are considered here. The total number of UCDs in the sample from \citet{sivakoff2007a} is about 400, as can be calculated from $f_{\rm LMXB}$ in the according intervals.

For practical purposes, it is useful not to discuss individual values for the $f_{\rm LMXB}$ of UCDs, but to replace them by a continuous function $\overline{P}(L_V)$. This function is obtained by performing a least-squares fit of a linear function to the values for $f_{\rm LMXB}$ in the $L_V$ intervals with the UCDs, leading to
\begin{equation}
\log_{10}(\overline{P})=a \log_{10}(L_V)+b,
\label{eq:P}
\end{equation}
where the best fitting parameters $a$ and $b$ are given in Tab.~(\ref{tab:parameter}). $\overline{P}$ can be interpreted as an estimate for the average probability for UCDs with a given $L_V$ to host a LMXB brighter than the detection limit. For a meaningful comparison between $\overline{P}$ and $\overline{\Gamma}_{\rm h}$ at different values for $L_V$, $A$ needs to be gauged. This is done by imposing that $\overline{P}(L_V)=\overline{\Gamma}_{\rm h}(L_V)$ for $L_V=10^6 \ L_{\odot,V}$. The motivation for choosing this condition to fix $A$ is that the stellar populations of systems with this luminosity should be nearly unaffected by dynamical evolution (cf. Section \ref{sec:properties}), while their $M/L$-ratios suggest that their IMF is canonical, in contrast to even more luminous stellar systems (cf. \citealt{dabringhausen2009a}).

If the rate at which LMXBs are produced in GCs and UCDs is proportional to some power $\gamma$ of the encounter rate in them, leading to $\overline{P}(L_V) \propto \overline{\Gamma}_{\rm h}^{\gamma}$ (cf. Equations~\ref{eq:prob} and~\ref{eq:P}), Equation~(\ref{eq:ratio2}) can be transformed into
\begin{equation}
\frac{\overline{P}(L_V)^{\frac{1}{\gamma}}}{\overline{\Gamma}_{\rm h}(L_V)}= \\
\frac{n_{\rm ns}(\alpha_3)}{n_{\rm ns}(\alpha_3=2.3)} \sqrt{\frac{M(\alpha_3=2.3)}{M(\alpha_3)}}.
\label{eq:ratio3}
\end{equation}
The left side of Equation~(\ref{eq:ratio3}) is then expressed in terms of observable properties of UCDs and the right side only depends on $\alpha_3$ as a free parameter once $m_{\rm tr}$ is given. Equation~(\ref{eq:ratio3}) can therefore be used to estimate the dependency of the IMF of the UCDs as a function of their observed $L_V$. Since $A$ is chosen such that $\overline{P}(L_V)/\overline{\Gamma}(L_V)=1$ for stellar systems that are assumed to have formed with the canonical IMF, $\overline{P}(L_V)/\overline{\Gamma}(L_V)>1$ implies a top-heavy IMF and $\overline{P}(L_V)/\overline{\Gamma}(L_V)<1$ implies a bottom-heavy IMF.

\subsection{Results}
\label{sec:results}

In order to test for an 
LMXB-excess and thus a top-heavy IMF in UCDs from the observational data from \citet{sivakoff2007a}
we now compare with theoretically 
expected LMXB-frequencies.
 
The dynamical formation of LMXBs depends on the density of both 
dark remnants and low-mass stars (Equation~\ref{eq:Gamma3}). In denser star clusters,
close encounters are more frequent and the formation of an LMXB is
more likely. GCs have a common half-mass radius
of a few parsec independent of their luminosity and their stellar mass is on average proportional to their luminosity \citep{mclaughlin2000a}. It therefore follows from Equation~(\ref{eq:GammaGC}) that LMXBs should be hosted
predominantly in high-mass GCs if their SRMF does not depend on their stellar mass. The dashed line
in Figure~\ref{lmbx-lv} shows the theoretically expected LMXB frequency 
for a constant IMF calculated with Equations~(\ref{eq:GammaGC}) and~(\ref{eq:GammaUCD}) with $A$ chosen such that these equations reproduce the observed LMXB frequency at $L_V=10^6 \ {\rm L}_{\odot}$. The theoretical prediction then matches the observations in the GC regime (i.e. $L_V < 10^6 \ {\rm L}_{\odot}$), in 
agreement with earlier studies on LMXBs in GCs \citep{jordan2005a,peacock2010a}.

At $L_V \approx 10^6 \ {\rm L}_{\odot}$, the transition luminosity from GCs to UCDs, 
both kinds of stellar systems have the 
same half-mass radius (see Section~\ref{sec:properties}). However, unlike GCs, UCDs show a luminosity-radius relation 
such that they become less dense with increasing
luminosity (cf. figure~4 in \citealt{dabringhausen2008a}). Consequently, Equation~(\ref{eq:GammaUCD}) predicts that the capture rate of late-type stars by 
dark remnants and thus the expected LMXB frequency decreases rapidly 
with increasing $L_V$-band luminosity if the SRMF is constant. Note that a constant SRMF in UCDs implies a constant IMF in them due to the lack of dynamical evolution in UCDs (see Section~\ref{sec:properties}). 

In Figure~\ref{lmbx-lv}, the prediction from Equation~(\ref{eq:GammaUCD}) for the LMXB frequency in UCDs with a constant SRMF is shown by the dashed line in the according luminosity range, where $A$ is chosen such that Equation~(\ref{eq:GammaUCD}) reproduces the observed LMXB frequency at $L_V=10^6 \ {\rm L}_{\odot}$. Two cases are considered, namely $f_{\rm LMXB} \propto \Gamma_{\rm h}$ and $f_{\rm LMXB} \propto \Gamma_{\rm h}^{0.8}$. The second case is closer to the dependency between $f_{\rm LMXB}$ and $\Gamma_{\rm h}$ reported by \citet{sivakoff2007a}. The agreement between the theoretical prediction and and the observed frequency of LMXBs is in either case good for GCs. However, $f_{\rm LMXB}\propto \Gamma_{\rm h}^{0.8}$ seems indeed a better fit to the data than $f_{\rm LMXB} \propto \Gamma_{\rm h}$. Note that \citet{maccarone2011a} find $\Gamma \propto \Gamma_{\rm h}^{0.8}$ on average for GCs in the Milky Way, which essentially means that the typical ratio between $\Gamma$ and $\Gamma_{\rm h}$ depends for these GCs on their mass.  This is probably a consequence of the more massive GCs in the MW being more concentrated than the less massive ones (\citealt{mclaughlin2000a}; cf Section~\ref{sec:constantIMF}), and likely to be the case for the GCs in the Virgo cluster as well.

For UCDs however, the observed LMXB-frequency strongly deviates from the theoretical prediction for a constant SRMF. It is observed that $25\pm5$ per cent of the UCDs with $L_V \approx 5 \times 10^6 \ L_{\odot,V}$ have a bright LMXB, while Equation~(\ref{eq:GammaUCD}) suggests a LMXB frequency of about 2 per cent at this luminosity for $f_{\rm LMXB} \propto \Gamma_{\rm h}$ and a LMXB frequency of about 3 per cent for $f_{\rm LMXB} \propto \Gamma_{\rm h}^{0.8}$. Thus, the expected fraction of LMXBs hosting UCDs is up to $\approx 10$ times smaller
than observed if all UCDs had the same IMF.

This discrepancy between the data and the model with an invariant (canonical) IMF and the
data is highly significant. This cannot be explained with more dark remnants remaining bound to UCDs due to higher escape velocities. This is because the escape velocity from massive GCs is much higher than the escape velocity from light GCs, since the characteristic radii of GCs do no change with mass, but the encounter rate is nevertheless sufficient for quantifying which fraction of them has a bright LMXB.

The situation is more complicated with the finding that redder GCs and UCDs have more LMXBs than the blue ones, while brighter objects (i.e. the UCDs in particular) tend to be redder than the less luminous ones \citep{mieske2006a}. Taking color as an indicator for metallicity leads to the interpretation that the LMXB-frequency in GCs and UCDs does not only depend on $\Gamma$ or $\Gamma_{\rm h}$ but also on metallicity \citep{jordan2004a,sivakoff2007a}. Note that an increase of metallicity with luminosity and therefore mass of GCs is consistent with theoretical modeling, according to which more massive star clusters retain more processed (i.e. metal-enriched) gas which is turned into subsequent stellar populations \citep{tenorio2003a}.

Using metallicity (i.e. color) as a second parameter besides $\Gamma_{\rm h}$ indeed allows a more precise modeling of the probability to find a LMXB in a given GC or UCD than when $\Gamma_{\rm h}$ is assumed to be the sole parameter determining the probability to find a LMXB in that GC or UCD \citep{sivakoff2007a}. The dependency of that probability is however nevertheless almost linear to the encounter rate, while the dependency on the metallicity is much weaker \citep{jordan2004a,sivakoff2007a}. This may explain why the fraction of GCs with a LMXB is apparently already well explained if only the encounter rate in the GCs is considered (see Figure~\ref{lmbx-lv}) despite the color-luminosity relation for GCs in the Virgo cluster (cf. \citealt{mieske2006a}). It is thereby unlikely that the drastic discrepancy between the observed  LMXB-frequency in UCDs and the theoretical prediction based on the encounter rate can be explained by an unaccounted metallicity effect, even though the color-luminosity dependency may be somewhat more pronounced for UCDs than for GCs \citep{mieske2010a}.

The conclusion is that the large number of LMXBs in UCDs is best explained by a large number dark remnants as a consequence of a top-heavy IMF in UCDs (and not as a consequence of different escape velocities or metallicities).

For an invariant IMF the theoretical LMXB frequency is highest at a
luminosity of $L_V \approx 10^6 \ L_{\odot,V}$, because in these systems
the present-day stellar density has a maximum and close encounters
are most frequent (Figure~4 in \citealt{dabringhausen2008a}). If the very dense star formation conditions
are responsible for a top-heavy IMF then, on first sight,
the smallest IMF index $\alpha_3$ is expected in systems with 
$L_V \approx 10^6 \ L_{\odot,V}$ and not in the most luminous UCDs. 
However, in systems with a top-heavy IMF stellar feedback is strongly enhanced
and rapid gas expulsion leads to an expansion of the 
UCDs \citep{dabringhausen2010a}. The UCDs revirialise after a few dynamical 
time scales ($\lesssim$100~Myr) and undergo no further size evolution. 
Thus, their present day stellar density is the dynamically relevant quantity
for producing the LMXB population.

We now determine by what amount the dark remnant content in UCDs 
has to be increased to get the theoretical LMXB-frequency 
into agreement with the observed values. For this, Equation~(\ref{eq:ratio3}) with $\gamma=1$ and $\gamma=0.8$ is used. This equation has $\alpha_3$ and $m_{\rm tr}$ as parameters (Equation~\ref{eq:IMF}). In this paper, $m_{\rm tr}=1 \ {\rm M}_{\odot}$ and $m_{\rm tr}=5 \ {\rm M}_{\odot}$, so that the influence of the in principle quite arbitrary choice of $m_{\rm tr}$ is tested. Note that with $m_{\rm tr}=1 \ {\rm M}_{\odot}$, Equation~(\ref{eq:IMF}) describes the family of IMFs that were considered in \citet{dabringhausen2009a}.  For either choice of $m_{\rm tr}$, the canonical IMF \citep{kroupa2001a,kroupa2002a} corresponds to $\alpha_3=2.3$ and a smaller value of $\alpha_3$ increases the fraction of high-mass
stars and subsequent dark remnants. The $\alpha_3$ that can explain the discrepancy between $\overline{P}(L_V)$ (i.e. the function describing the observed LMXB frequency in UCDs) and $\overline{\Gamma}_{\rm h}(L_V)$ (i.e. the theoretical expectation for the LMXB frequency in UCDs if their IMF was canonical) at a given $L_V$ can be found by numerically solving Equation~(\ref{eq:ratio3}) for $\alpha_3$ with a given value for $m_{\rm tr}$.

\begin{figure*}[t]
\centering
  \includegraphics[scale=0.7]{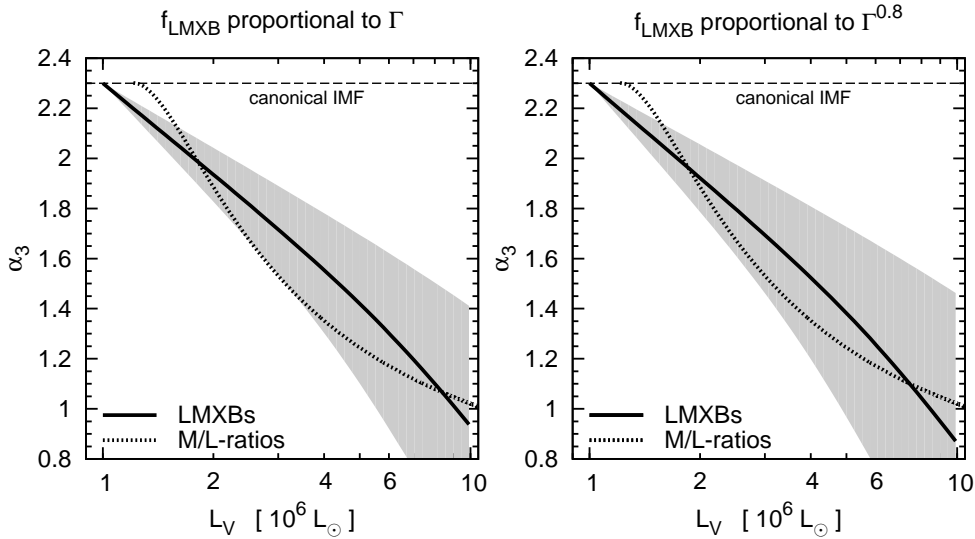}
  \caption{\label{alpha} The IMF in UCDs for $m_{\rm tr} = 1 \ {\rm M}_{\odot}$.
  Plotted is the high-mass IMF index, $\alpha_3$, as a function
  of the V-band luminosity of the UCDs, $L_V$ for $f_{\rm LMXB} \propto \Gamma_{\rm h}$ leading to $P(L_V) \propto \overline{\Gamma}_{\rm h}$ (left panel) and for $f_{\rm LMXB} \propto \Gamma_{\rm h}^{0.8}$ leading to $P(L_V) \propto \overline{\Gamma}_{\rm h}^{0.8}$ (right panel).
  The solid line shows
  the most likely high-mass index required to increase the
  dark remnant content in UCDs in order to match to observed  
  LMXB-frequency (derived from the 
  solid line in Figure~\ref{lmbx-lv}). The grey shaded area
  marks an estimate for the 3~$\sigma$ region. The horizontal long dashed line
  marks the canonical IMF with $\alpha_3 = 2.3$.
  The dotted line shows the independently 
  calculated high-mass IMF index obtained from the observed 
  mass-to-light ratios of UCDs \citep{dabringhausen2009a}.
  Simple-to-use fitting relations for the 
  variation of $\alpha_3$ with $L_V$ can be found in Tab.~(\ref{tab:parameter}). 
  }
\end{figure*}

\begin{figure*}[t]
\centering
  \includegraphics[scale=0.7]{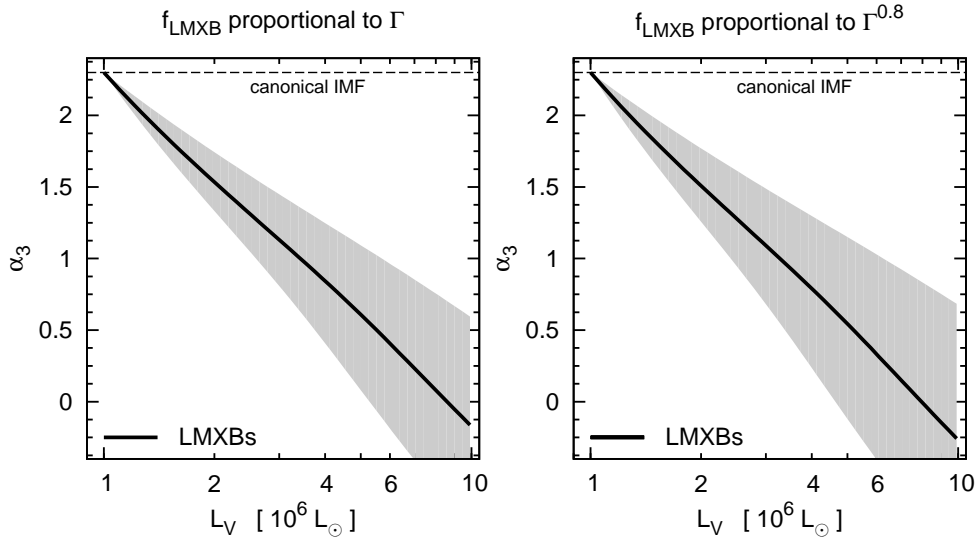}
  \caption{\label{alpha5} The IMF in UCDs for $m_{\rm tr} = 5 \ {\rm M}_{\odot}$, otherwise as Figure~\ref{alpha}.
  }
\end{figure*}

The $L_V$ dependence of $\alpha_3$ required to bring the model into agreement with 
the UCD data is plotted as the solid line in Figure~\ref{alpha} for $m_{\rm tr} =1 \ {\rm M}_{\odot}$ and in Fig~(\ref{alpha5}) for $m_{\rm tr} = 5 \ {\rm M}_{\odot}$. In either case, the most massive UCDs must have an extremely 
top-heavy IMF in order to explain their LMXB-excess. The higher $m_{\rm tr}$ is, the more exotic the IMF of UCDs must be in order to explain the number of LMXBs in them. For a given value for $m_{\rm tr}$, it is on the other hand only of minor importance whether $P(L_V)$ is proprotional to $\overline{\Gamma}_{\rm h}$ or proportional to $\overline{\Gamma}_{\rm h}^{0.8}$.

For $m_{\rm tr}=1 \ {\rm M}_{\odot}$, the
independent analysis in this paper leads the same top-heavy IMF as derived from the UCD mass-to-light 
ratios \citep{dabringhausen2009a}, shown as
the dotted line in Figure~\ref{alpha}. Such a comparison is not meaningful for $m_{\rm tr}= 5 \ {\rm M}_{\odot}$, since the shape thereby assumed for the IMF is different from the IMF considered in \citet{dabringhausen2009a}.

The most likely relations between $\alpha_3$ and $\log_{10}(L_V)$ shown in Figures~\ref{alpha} and~\ref{alpha5} are remarkably close to a linear function, 
\begin{equation}
\overline{\alpha_3}=c \log_{10}(L_V)+d.
\end{equation}
The best fitting parameters $c$ and $d$ have been determined from a least-squares fit to 48 sample values calculated from Equation~(\ref{eq:ratio3}). These are shown Table~\ref{tab:parameter}. Probably the best model for the IMF in UCDs is calculated when $f_{\rm LMXB} \propto \Gamma_{\rm h}^{0.8}$ and $m_{\rm tr} = 1 \ {\rm M}_{\odot}$ are assumed. This is because observations suggest a less-than-linear dependency of $f_{\rm LMXB}$ on the encounter rate \citep{jordan2004a,sivakoff2007a} and assuming $m_{\rm tr} > 1 \ {\rm M}_{\odot}$ implies even more extreme deviations from the canonical IMF in high-mass UCDs while the IMF is remarkably invariant in open star clusters \citep{kroupa2001a}.

Figure~\ref{lmbx-lv} suggests that the value of $f_{\rm LMXB}$ for the most luminous UCDs is of central importance for estimating the slope of $\overline{P}(L_V)$ (Equation~\ref{eq:P}) and thus for the $\alpha_3$ calculated from Equation~(\ref{eq:ratio3}). This is because of the distance of these data points to the other data points, which is due to the fact that the corresponding $L_V$ interval is large. In order to estimate an uncertainty to the dependency of $\alpha_3$ on $L_V$, we changed the value of $f_{\rm LMXB}$ for the most luminous UCDs ($L_V \gtrsim 2 \times 10^6 \ {\rm L}_{\odot}$) by 3 times its uncertainty. $\overline{P}(L_V)$ was then recalculated with this new value and used in Equation~(\ref{eq:ratio3}). The resulting limits on the dependency of $\alpha_3$ on $L_V$ are indicated by the limits to the gray area in Figures~\ref{alpha} and~\ref{alpha5}. Also the limits of the gray areas are parametrized with linear functions, which are listed in Table~\ref{tab:parameter}.

\begin{figure*}[t]
\centering
\includegraphics[scale=0.7]{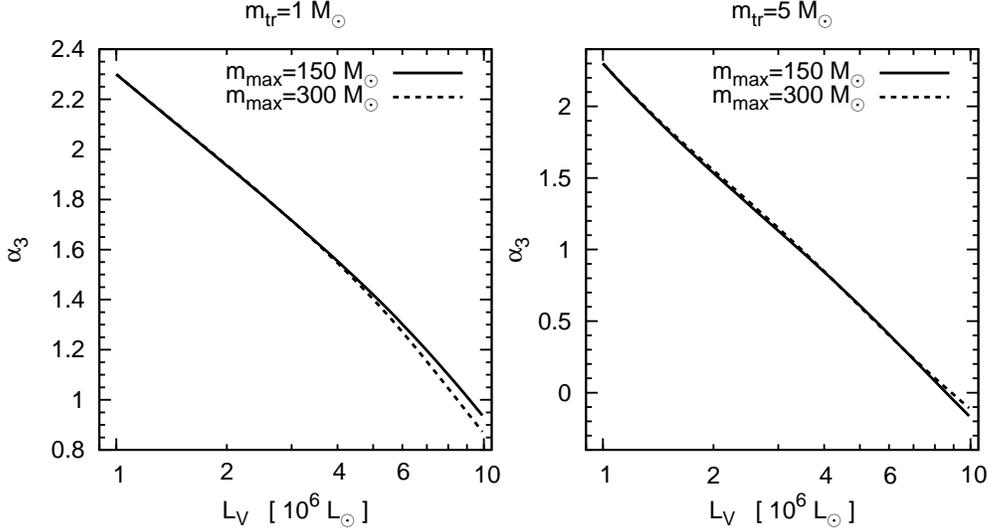}
\caption{\label{fig:comp} The high-mass IMF slope, $\alpha_3$, as a function of $V$-band luminosity for different upper mass limits of the IMF, $m_{\rm max *}$ assuming $m_{\rm tr} = 1 \ {\rm M}_{\odot}$ (left panel) or $m_{\rm tr} = 5 \ {\rm M}_{\odot}$ (right panel). The solutions, $\alpha_3(L_V)$, are calculated from Equation~(\ref{eq:ratio3}) with $\gamma=1$, i.e under the assumption that $f_{\rm LMXB} \propto \Gamma_{\rm h}$. Assuming $\gamma=0.8$ leads however qualitatively to the same results as assuming $\gamma=1$ (c.f. Figures~\ref{alpha} and~\ref{alpha5}). The solid line corresponds in both panels to $m_{\rm max *}=150 \  {\rm M}_{\odot}$ and is thus identical with the solid line in Figure~\ref{alpha}, and Figure~\ref{alpha5}, respectively. The dashed lines corresponds to $m_{\rm max *}=300 \  {\rm M}_{\odot}$.}
\end{figure*}

The uncertainty of the upper mass limit for stars, $m_{\rm max *}$, has little effect on the results summarized in Table~\ref{tab:parameter}. This is illustrated with Figure~\ref{fig:comp}, where the dependency between $L_V$ and $\alpha_3$ calculated from Equation~(\ref{eq:ratio3}) is shown for $m_{\rm max *}=150 \ {\rm M}_{\odot}$ \citep{weidner2004a,oey2005a} and for $m_{\rm max *}=300 \ {\rm M}_{\odot}$ \citep{crowther2010a}. The two functions are almost identical.

\begin{table*}
\caption{\label{tab:parameter} The best fitting parameters for linear fits to $P$ and $\alpha_3$ for different models.  The different cases (most likely case, upper limit, lower limit) listed here for every model correspond to different values of $P$ for the UCDs with the highest masses (cf. Sec~\ref{sec:results}). Probably the best model for the IMF in UCDs is calculated when $f_{\rm LMXB} \propto \Gamma_{\rm h}^{0.8}$ and $m_{\rm tr} = 1 \ {\rm M}_{\odot}$ are assumed. This is because observations suggest a less-than-linear dependency of $f_{\rm LMXB}$ on the encounter rate \citep{jordan2004a,sivakoff2007a} and assuming $m_{\rm tr} > 1 \ {\rm M}_{\odot}$ implies even more extreme deviations from the canonical IMF in high-mass UCDs while the IMF is remarkably invariant in open star clusters \citep{kroupa2001a}. The parameters describing the IMF according to this model are shown in bold face in this table.}
\centering
\vspace{2mm}
\begin{tabular}{lrrrrr}
\hline
&&&&\\[-10pt]
&\multicolumn{2}{c}{$\log_{10}(\overline{P})=a \log_{10}(L_V)+b$} & & \multicolumn{2}{c}{$\overline{\alpha_3}=c \log_{10}(L_V)+d$}\\
model 	   & $a$ 	     & $b$ 	   &   & $c$ 	     & $d$ \\
\hline
$f_{\rm LMXB} \propto \Gamma_{\rm h}$, $m_{\rm tr} = 1 \ {\rm M}_{\odot}$, most likely case & $0.207$ & $1.249$ & & $-1.337$ & $2.332$ \\
$f_{\rm LMXB} \propto \Gamma_{\rm h}$, $m_{\rm tr} = 1 \ {\rm M}_{\odot}$, upper limit & $0.615$ & $1.201$ & & $-1.878$ & $2.375$ \\
$f_{\rm LMXB} \propto \Gamma_{\rm h}$, $m_{\rm tr} = 1 \ {\rm M}_{\odot}$, lower limit & $-0.202$ & $1.298$ & & $-0.884$ & $2.396$ \\
\hline
$f_{\rm LMXB} \propto \Gamma_{\rm h}^{0.8}$, $m_{\rm tr} = 1 \ {\rm M}_{\odot}$, most likely case & $0.207$ & $1.249$ & & ${\bf -1.402}$ & ${\bf 2.337}$ \\
$f_{\rm LMXB} \propto \Gamma_{\rm h}^{0.8}$, $m_{\rm tr} = 1 \ {\rm M}_{\odot}$, upper limit & $0.615$ & $1.201$ & & ${\bf -2.089}$ & ${\bf 2.391}$ \\
$f_{\rm LMXB} \propto \Gamma_{\rm h}^{0.8}$, $m_{\rm tr} = 1 \ {\rm M}_{\odot}$, lower limit & $-0.202$ & $1.298$ & & ${\bf -0.861}$ & ${\bf 2.304}$ \\
\hline
$f_{\rm LMXB} \propto \Gamma_{\rm h}$, $m_{\rm tr} = 5 \ {\rm M}_{\odot}$, most likely case & $0.207$ & $1.249$ & & $-2.415$ & $2.275$ \\
$f_{\rm LMXB} \propto \Gamma_{\rm h}$, $m_{\rm tr} = 5 \ {\rm M}_{\odot}$, upper limit & $0.615$ & $1.201$ & & $-3.169$ & $2.289$ \\
$f_{\rm LMXB} \propto \Gamma_{\rm h}$, $m_{\rm tr} = 5 \ {\rm M}_{\odot}$, lower limit & $-0.202$ & $1.298$ & & $-1.679$ & $2.263$ \\
\hline
$f_{\rm LMXB} \propto \Gamma_{\rm h}^{0.8}$, $m_{\rm tr} = 5 \ {\rm M}_{\odot}$, most likely case & $0.207$ & $1.249$ & & $-2.512$ & $2.277$ \\
$f_{\rm LMXB} \propto \Gamma_{\rm h}^{0.8}$, $m_{\rm tr} = 5 \ {\rm M}_{\odot}$, upper limit & $0.615$ & $1.201$ & & $-3.442$ & $2.290$ \\
$f_{\rm LMXB} \propto \Gamma_{\rm h}^{0.8}$, $m_{\rm tr} = 5 \ {\rm M}_{\odot}$, lower limit & $-0.202$ & $1.298$ & & $-1.594$ & $2.263$ \\

&&&&\\[-10pt]
\hline
\end{tabular}
\end{table*}

\section{The supernova rate in Arp~220}
\label{sec:SNrate}

A top-heavy IMF in UCDs can theoretically be understood if UCDs formed as very massive star clusters that were internally heated by infra-red radiation that was trapped inside a molecular cloud massive enough to form a UCD-type star cluster \citep{murray2009a}, or if UCDs formed from molecular clouds that were heated by highly energetic cosmic rays originating from numerous type-II supernovae surrounding those molecular clouds (\citealt{papadopoulos2010a}; cf. Section~\ref{sec:intro}). Both scenarios imply that UCDs are formed during star-bursts, either because of the link between the formation of the most massive star-clusters and high star formation rates \citep{weidner2004b}, or because the cosmic-ray field would only then be intense enough for effective heating of the molecular clouds. Note that likely progenitors of UCDs have actually been observed in star-bursts \citep{fellhauer2002a}.

Ultra-luminous infra-red galaxies (ULIRGs) are believed
to be galaxies with star-bursting regions \citep{condon1991a}. They are thus systems where UCDs are probably forming. If this notion is correct and the IMF in UCDs is top-heavy, the ULIRGs as a whole should have more massive stars than expected for an invariant, canonical IMF. As a consequence, the rate of type~II supernovae is expected to be higher.

In the following, we test the hypothesis of a top-heavy IMF in ULIRGs. For this reason, we quantify how the type-II supernova rate (SNR) in a star burst is connected to the star formation in it. Based on this, theoretical predictions for the SNR of Arp~220, which is one of the closest ULIRGs \citep{lonsdale2006a}, are calculated and compared to observations of this stellar system.

The type-II supernova rate (SNR) observed in a stellar system depends on its IMF as well as on its star formation history (SFH), i.e. how the star formation rate in the stellar system has changed with time, because these quantities determine the numbers and ages of stars in given mass intervals. If star formation begins at a time $t_0$, only stars above a time-dependent mass-limit $m_{\rm low}$ can have completed their evolution at a time $t>t_0$. For stars evolving into SNe, this mass can be approximated \citep{dabringhausen2010a} by
\begin{equation}
\frac{m_{\rm low}}{{\rm M}_{\odot}}=74.6\left(\frac{t-t_0}{\rm Myr}-2.59\right)^{-0.63}.
\label{eq:lifetimes}
\end{equation}
Note that no stars evolve to type-II supernovae (SNe), if $t-t_0\le 2.59 \ {\rm Myr}$.

Now consider a time interval $[t, t+\Delta t]$ and stars in a mass interval $[m, m+\Delta m]$, where $m\ge m_{\rm low}$. If the SFR was constant for all $t\ge t_0$, the number of stars evolving into SNe in the given mass interval during the time $\Delta t$ is equal to the number of new stars  that are formed in the same mass interval. Thus,
\begin{equation}
\frac{\Delta {\rm SNR}}{\rm yr^{-1}}=\frac{\rm SFR}{{\rm M}_{\odot} \ {\rm yr}^{-1}}\int^{m+\Delta m}_{m} \xi (m) \, dm
\label{eq:SNR1}
\end{equation} 
in this case, where $\xi (m)$ is assumed to be given by Equation~(\ref{eq:IMF}) with the normalization defined by Equation~(\ref{eq:norm2}). This normalization keeps the total mass of the stars which are formed per unit time constant.

If $\Delta t$ is small compared to the time scale on which $m_{\rm low}$ changes, the number of all stars that evolve during $\Delta t$ can be approximated as
\begin{equation}
\frac{\rm SNR}{\rm yr^{-1}} \approx \frac{\rm SFR}{{\rm M}_{\odot} \ {\rm yr}^{-1}}\int^{m_{\rm max *}}_{m_{\rm low}} \xi (m) \, dm.
\label{eq:SNR2}
\end{equation}
Note that the SFR in Equations~(\ref{eq:SNR1}) and~(\ref{eq:SNR2}) should be considered an average value over a time-scale $t-t_0$. Variations of the SFR on much shorter time-scales are of no importance here.

The SFR of a ultra-luminous infra-red galaxy (ULIRG) can be estimated as
\begin{equation}
\frac{\rm SFR}{{\rm M}_{\odot} \ {\rm yr}^{-1}}=\frac{L_{\rm FIR}}{5.8 \times 10^9 \ {\rm L}_{\odot}},
\label{eq:SFR}
\end{equation}
where $L_{\rm FIR}$ is the far infra-red (FIR) luminosity of the ULIRG \citep{kennicutt1998a}.

One of the nearest ULIRGs is Arp~220. Using $L_{\rm FIR}=1.41\times10^{12} \ {\rm L}_{\odot}$ for Arp~220 \citep{sanders2003a}, Equation~(\ref{eq:SFR}) implies a SFR of $\approx 240 \ {\rm M}_{\odot} \, {\rm yr}^{-1}$ for that galaxy. The SNe in Arp~220 have been observed in a central region with a diameter of $\approx 1\ {\rm kpc}$, from where about 40 per cent of its FIR luminosity originates \citep{soifer1999a}. Equation~(\ref{eq:SFR}) thus implies a SFR of $\approx 100 \ {\rm M}_{\odot} \, {\rm yr}^{-1}$ if only this part of Arp~220 is considered. Note that this SFR is consistent with the SFR that has been suggested for a forming UCD if UCDs form on a timescale of approximately 1 Myr \citep{dabringhausen2009a}. Also note that the observed SN in Arp~220 do not seem to distributed evenly over the central part of Arp~220, but to be concentrated in two knots which have a radius $\approx$ 50 pc each \citep{lonsdale2006a} (i.e. the size of a UCD). This implies that indeed a major part of the star formation in the central part of Arp~220 takes place within these two knots. This would imply projected star formation densities of a few $10^{-3} \ {\rm M}_{\odot} \ {\rm yr}^{-1} \ {\rm pc}^{-2}$ in the knots.

\begin{figure}[t]
\centering
\includegraphics[scale=0.7]{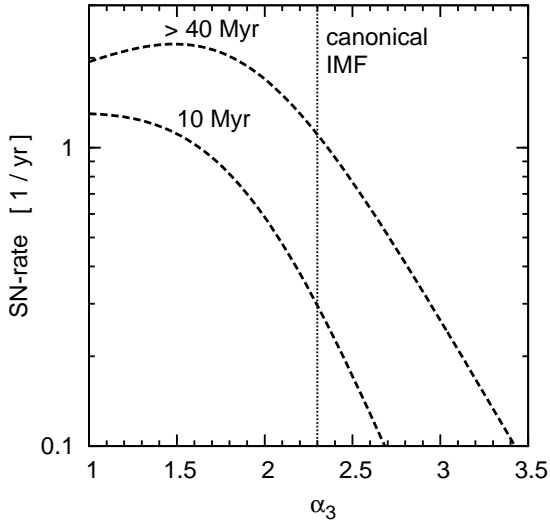}
\caption{\label{fig:SNrate}The SN-rate in the center of Arp~220. The SN-rates are functions of the slope of the IMF above a stellar mass of $1 \ {\rm M}_{\odot}$, $\alpha_3$, but also depend on the length of the star burst (which is indicated above the corresponding curve). They do not exceed $\approx 1$ SN~yr$^{-1}$ for the canonical IMF (whose high-mass slope is marked by the dotted vertical line), or $\approx 2$ SN~yr$^{-1}$ for a top-heavy IMF.}
\end{figure}

SNRs calculated from Equation~(\ref{eq:SNR2}) for a constant SFR of $100 \ {\rm M}_{\odot} \, {\rm yr}^{-1}$ are shown as functions of the high-mass slope of the IMF in Figure~\ref{fig:SNrate}. The two curves correspond to different times at which the star burst was initialized, but the expected number of SN per year (i.e. the SNR) is low in any case. The number of SN that \emph{actually} occur within one year can therefore differ substantially from the calculated SNR, as the frequency of SN over such a time span obeys low-number statistics. Thus, the probability for a certain number of SN to happen within one year is quantified by the Poisson distribution function. 

Now consider the case that the star-burst in Arp~220 already lasts for more than 40~Myr. This implies $m_{\rm low} = 8 \ {\rm M}_{\odot}$, so that the number of SNII per year is maximized for the given SFR. The expectation value for the SNII-rate is then about one per year if the IMF was canonical (i.e. $\alpha_3=2.3$), but about two per year for a top-heavy IMF with $1 \lesssim \alpha_3 \lesssim 2$, where the SN-rate is only a weak function of $\alpha_3$ (cf. Figure~\ref{fig:SNrate}). Thus, the probability to actually observe four new SNII in a given year \citep{lonsdale2006a} is then about two per cent if the IMF is canonical, but about 12 per cent for $1 \lesssim \alpha_3 \lesssim 2$.

A more elaborate discussion of the SNII rate in Arp~220 is obtained by taking into account that stars in a galaxy form in star-clusters of different masses, since $m_{\rm max}$ of the IMF depends on the mass of the star-cluster for low-mass star-clusters. This implies that the integrated galactic IMF (IGIMF) of all star-clusters in Arp~220 combined is not equal to the IMF in its star-clusters.

This IGIMF is given by
\begin{eqnarray}
\xi_{\rm IGIMF}(m) & = & \int^{M_{\rm ecl,max}{\rm (SFR)}}_{M_{\rm ecl,min}} \xi (m \le m_{\rm max}(M_{\rm ecl})) \nonumber \\
& \times & \xi_{\rm ecl} (M_{\rm ecl}) \, dM_{\rm ecl},
\end{eqnarray}
where $m$ is the initial stellar mass, $M_{\rm ecl}$ is the initial stellar mass of a star cluster, $M_{\rm ecl,min}$ is the minimum mass of newly formed star-clusters, $M_{\rm ecl,max}{\rm (SFR)}$ is the SFR-dependent maximum mass of newly formed star-clusters, $\xi (m)$ is the IMF and $\xi_{\rm ecl}(M_{\rm ecl})$ is the star-cluster mass function \citep{weidner2005a,weidner2010a}. The IGIMF can be parametrized by a multi-power law,
\begin{equation}
\xi_{\rm IGIMF} (m) =k k_i m^{-\alpha_i},
\label{eq:IGIMF}
\end{equation}
with 
\begin{eqnarray}
\nonumber \alpha_1 = 1.3,  & \qquad & 0.1  \le  \frac{m}{\rm{M}_{\odot}} < 0.5,\\
\nonumber \alpha_2 = 2.3,  & \qquad & 0.5  \le  \frac{m}{\rm{M}_{\odot}} < 1, \\
\nonumber \alpha_{\rm IGIMF} \, \in \, \mathbb{R}, & \qquad & 1 \le  \frac{m}{{\rm M}_{\odot}} \le m_{\rm max *},
\end{eqnarray}
where the factors $k_i$ ensure that the IGIMF is continuous where the power changes and $k$ is a normalization constant. $\xi_{\rm IGIMF}(m)$ equals 0 if $m<0.1 \, {\rm M}_{\odot}$ or $m> m_{\rm max *}$,  where $m_{\rm max *}$ is the maximum stellar mass. Thus, the IGIMF defined here is equal to the IMF defined by Equation~(\ref{eq:IMF}), except for the high-mass slope and the upper mass limit. 

The case of a canonical IMF in all star-clusters, i.e. $\alpha_3=2.3$, implies \citep{weidner2005a} $\alpha_{\rm IGIMF} \gtrsim 3$. The expectation value for the number of SNII per year would then be $\lesssim 0.2$ per year. On the other hand, $\alpha_{\rm IGIMF} \lesssim 2$ is possible, if a varying IMF that becomes more top-heavy with star-cluster mass is considered \citep{weidner2010a,kroupa2012a}. This implies that the probability to actually observe four new SNII in a given year \citep{lonsdale2006a} is essentially zero if the IMF is canonical in all star-clusters, but it can still be about 10 per cent if the IMF becomes top-heavy in massive UCD-type star-clusters.

The remnants produced by SNII are neutron stars and black holes. The SNII-rates thereby are an indicator for how many mergers of such remnants can be detected by searching for gravitational waves. Comparing the SN-rate for $\alpha_{\rm IGIMF}=3$ to the SN-rate for $\alpha_{\rm IGIMF}=2$ thus suggests that about an order of magnitude more of such events may be expected if the IMF in massive star-clusters is not canonical, but top-heavy. Thus, the hitherto predicted detection rate of about 30 mergers of dark remnants per year \citep{banerjee2010a} for the upcoming adLIGO-experiment could be too low by an order of magnitude, as an invariant IMF has been used for this estimate.

Further evidence for a top-heavy IMF in star-bursting galaxies is found by \citet{anderson2011a} in Arp~299. They study numbers of different types of supernovae in Arp~299 and conclude from the mass of the appropriate progenitor stars that the IMF is probably top-heavy in that system. Thus, \citet{anderson2011a} qualitatively come to the same conclusion for Arp~299 as we did for Arp~220, while their method is different.

\section{Star formation densities and the IMF}
\label{sec:discussion}

A top-heavy IMF in UCDs is in-line with different 
studies concluding a top-heavy IMF in high-redshift star forming
galaxies \citep{vandokkum2008a,loewenstein2006a}. Contrary to this, a recent
spectroscopic study of two low-redshift very massive elliptical 
galaxies suggests a hitherto unseen large population of low-mass
stars \citep{vandokkum2010a}, which has been predicted as a possible consequence
of cooling flows on massive ellipticals \citep{kroupa1994a}.
It is on the other hand unlikely that the majority of the UCDs formed in potential wells deeps enough to cause cooling flows.

Also note that the current stellar densities suggest that the star-formation densities (SFDs), i.e. the SFR per volume, of UCDs were very different from the SFDs of elliptical galaxies. Consider for instance an exemplar present-day UCD with $M=10^7 \, {\rm M}_{\odot}$ and $r_{\rm h}=10 \ {\rm pc}$ and an exemplar present-day elliptical galaxy with $M=10^{12} \ {\rm M}_{\odot}$ and $r_{\rm h}=10^4 \ {\rm pc}$. These values can be considered representative for typical UCDs and massive elliptical galaxies, respectively (cf. figure~4 in \citealt{misgeld2011a}). Star formation is thought to have proceeded quickly in UCDs and massive elliptical galaxies, so that the stellar population of the exemplar UCD may have formed within $10^7$~yr \citep{dabringhausen2009a} and the stellar population of the exemplar elliptical galaxy may have formed within $10^9$~yr \citep{thomas2005a}. This leads to a SFR of $1 \ {\rm M}_{\odot} \, {\rm yr}^{-1}$ for the exemplar UCD and to a SFR of $10^3 \ {\rm M}_{\odot} \, {\rm yr}^{-1}$ for the exemplar elliptical galaxy. The SFD can be estimated by dividing the SFR by $r_{\rm h}^3$, leading to a SFD of $10^{-3} \ {\rm M}_{\odot} \, {\rm yr}^{-1} \, {\rm pc}^{-3}$ for the exemplar UCD and a SFD of $10^{-9} \ {\rm M}_{\odot} \, {\rm yr}^{-1} \, {\rm pc}^{-3}$ for the exemplar elliptical galaxy. However, according to \citet{dabringhausen2010a} UCDs must have been even more compact when they formed ($r_{\rm h} \approx 1$ pc), since the mass loss following star formation with a top-heavy IMF must have expanded them to their present-day radii. With the masses of UCDs being $10^6 \ {\rm M}_{\odot} \lesssim M \lesssim 10^8 \ {\rm M}_{\odot}$, their SFRs ranged from $0.1 \ {\rm M}_{\odot} \, {\rm yr}^{-1}$ to $10 \ {\rm M}_{\odot} \, {\rm yr}^{-1}$ if they formed within 10 Myr. An initial $r_{\rm h}$ of 1 pc thereby implies SFDs ranging from  $0.1 \ {\rm M}_{\odot} \, {\rm yr}^{-1} \, {\rm pc}^{-3}$ to  $10 \ {\rm M}_{\odot} \, {\rm yr}^{-1} \, {\rm pc}^{-3}$. Thus, the SFDs of UCDs can easily be higher by six to ten orders of magnitude than the SFDs of massive elliptical galaxies. Massive ellipticals can therefore not serve as a proxy for the 
stellar population in UCDs.

It is therefore perhaps the SFD that determines whether the IMF in some region of space becomes top-heavy, and not the overall SFR in a forming stellar system. This is actually consistent with models why the IMF may become top-heavy: \citet{dabringhausen2010a} argue that the central densities in forming UCDs were so high ($\rho > 10^5 \ {\rm M}_{\odot} \, pc^{-3}$) that collisions and perhaps mergers between pre-stellar cores were important in them, in contrast to less massive stellar systems. Likewise, if the heating of molecular clouds by cosmic rays is the process by which the IMF becomes top-heavy \citep{papadopoulos2010a}, it is again not the number, but the number \emph{density} of the surrounding massive stars that makes heating of the molecular cloud effective.

\section{Conclusion}
\label{sec:conclusion}

The dynamical mass-to-light ratios of ultra compact dwarf galaxies (UCDs) are surprisingly high \citep{hasegan2005a,dabringhausen2008a,mieske2008a}. This finding was explained by \citet{dabringhausen2009a} with an IMF that has more massive stars than the canonical IMF deduced by \citet{kroupa2001a} from resolved stellar populations in the Milky Way. The high mass-to-light ratio of UCDs is then a consequence of a large population of dark remnants (i.e. neutron stars and black holes) in them.

These dark remnants become visible as X-ray sources if they accrete matter from a low-mass companion star. The rate at which low-mass X-ray binaries (LMXBs) are formed in globular clusters and UCDs scales with the number density of dark remnants (see Section~\ref{sec:origin}). Data on the fraction of UCDs that harbour a bright X-ray source \citep{sivakoff2007a} can therefore be used to confirm the presence of a large population of dark remnants in UCDs by a method that does not rely on the fact that dark remnants only increase the mass of a UCD, but not its luminosity. It is shown in this paper that LMXBs in UCDs are indeed up to 10 times more frequent than expected for an invariant, canonical IMF. The overabundance of LMXBs is used to quantify the dependence of the high-mass IMF-slope, $\alpha_3$, on the luminosity of UCDs. This function is essentially equal to the dependence between the luminosity of the UCDs and their $\alpha_3$ suggested in \citet{dabringhausen2009a} based on the mass-to-light ratios of UCDs (see Section~\ref{sec:results}). Note that the $L_V$ of present-day GCs and UCDs is just one of many properties of such systems. Dependencies of $\alpha_3$ on their initial mass,  initial density and their metallitcity are therefore discussed in \citet{marks2012a}.

UCDs can be understood as the most massive star-clusters which only form at extremely high galaxy-wide star formation rates  (SFRs) \citep{weidner2004a}. Alternatively, UCDs could form by the merger of gravitationally bound systems of star clusters as they are observed in interacting galaxies \citep{fellhauer2002a}. In either case, the formation of UCDs would be connected to star-bursts. Given that ultra-luminous infra-red galaxies (ULIRGs) are interpreted as galaxies with star-bursting regions \citep{condon1991a}, they should show indications of a top-heavy IMF as a consequence. The nearest ULIRG is Arp~220. We show that the observed rate of type~II supernovae in this ULIRG is indeed highly improbable if the IMF is invariant, but not if the IMF is top-heavy (see Section \ref{sec:SNrate}).

There are thus three mutually consistent arguments for a top-heavy IMF in UCDs or more generally star-bursting systems. Together with the evidence for the formation of UCDs being connected to star bursts, these arguments imply that the IMF becomes top-heavy in star-bursts (cf. \citealt{weidner2010a}). This finding stands in contrast to the prevalent notion that the IMF is invariant \citep{kroupa2001a,kroupa2002a,bastian2010a,kroupa2012a} and thereby has important implications. For instance, estimates of the SFR of a galaxy based on observations that are sensitive only to high-mass stars and the assumption of an invariant IMF (like Equation~\ref{eq:SFR}) are too high if the IMF actually is top-heavy. Consequently, estimates for the time scale on which the population of low-mass star in that galaxy is built up until the gas of the galaxy is depleted become too short. Also the chemical evolution of galaxies is different if the IMF in them can become top-heavy, since the nuclear reactions that occur in a star mainly depend on its mass. This has implications on their content of metals and planetary systems \citep{ghezzi2010a}. Furthermore, as more dark remnants are formed if the IMF is top-heavy, more dark-remnant mergers and thus gravitational-wave emitters should be detected in this case. Finally, the dynamical evolution of star clusters critically depends on the shape of the IMF \citep{dabringhausen2010a}.

\section*{Acknowledgments} 
J.D acknowledges support through DFG-grant KR1635/13 
and thanks ESO for financial support via a grant from the 
Director General Discretionary Fund in 2009. The authors wish to thank Tom Maccarone for some useful comments.

\bibliographystyle{aa}
\bibliography{new}

\end{document}